# Kontextbasierte Aktivitätserkennung – Synergie von Mensch und Technik in der Social Networked Industry

**Friedrich Niemann M.Sc.\*, Dr.-Ing. Christopher Reining**

Lehrstuhl für Förder- und Lagerwesen, Technische Universität Dortmund, Joseph-von-Fraunhofer-Str. 2-4, 44227 Dortmund, Deutschland, Tel. 0231 755 2765
\* Kontaktperson: friedrich.niemann@tu-dortmund.de

**Abstract:** In der Social Networked Industry steht die Zusammenarbeit von Mensch und Technik im Vordergrund. Grundvoraussetzung für eine synergetische Zusammenarbeit aller Akteure ist die Kommunikation, welche neben verbalen auch nonverbale Interaktionen umfasst. Um eine nonverbale Interaktion zu ermöglichen, müssen Maschinen in der Lage sein, menschliche Bewegungen zu erfassen und zu verstehen. Dieser Beitrag stellt die laufende Grundlagenforschung zur Analyse menschlicher Bewegungen mittels sensorgestützter Aktivitätserkennung vor und zeigt Anknüpfungspunkte für einen Transfer in industrielle Anwendungen. Im Fokus steht die Praxistauglichkeit der Aktivitätserkennung durch die Hinzunahme weiterer Datenströme wie beispielsweise den Positionsdaten logistischer Objekte und Hilfsmitteln, d. h. dem Kontext, in dem eine gewisse Aktivität ausgeführt wird.

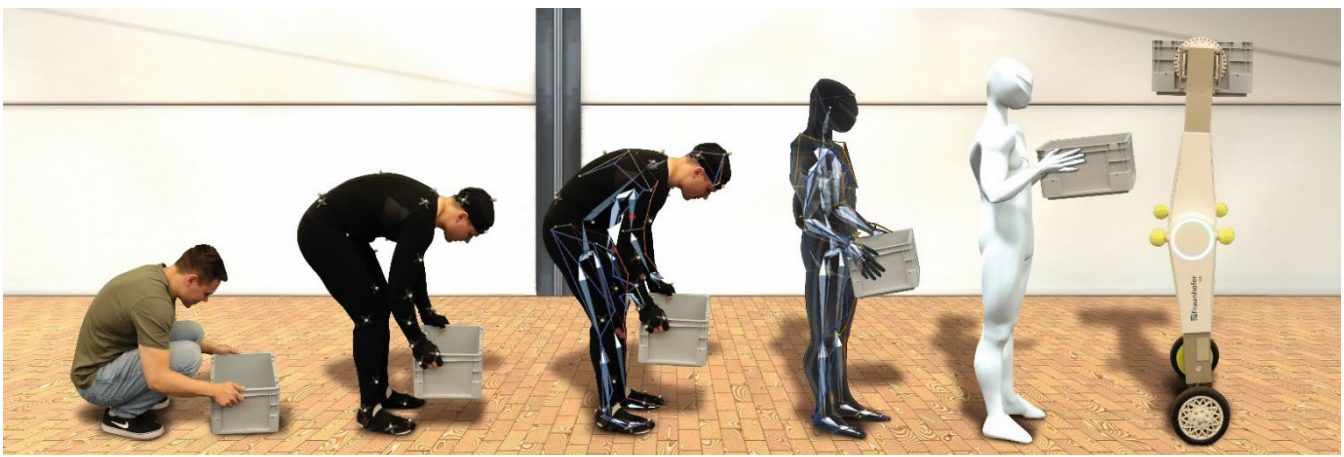

## 1. Effizienzsteigerung manueller Prozesse mittels automatischer Aktivitätserkennung

In der Social Networked Industry findet eine Verteilung von Aufgaben und Verantwortlichkeiten zwischen Mensch und Technik statt. Treiber der Veränderungen menschlicher Aktivitäten in der Intralogistik sind u. a. der Einzug intelligenter Sensoren sowie der zunehmende Verzicht auf starre Infrastrukturen zugunsten flexibler, modularer Materialflusssysteme. Trotz der damit einhergehenden Zunahme automatisierter Prozesse gehen Industrie und Forschung von einem steigenden Personalbedarf bei zugleich steigender Komplexität der Arbeitsaufgaben aus. Denn eine vollständige Automatisierung wird in naher Zukunft in den meisten Fällen weder technologisch noch wirtschaftlich sinnvoll sein. Manuelle (Teil-)Systeme lassen im Hinblick auf heterogene und häufig wechselnde Produktportfolios sowie schwankende Durchsatzanforderungen mehr Wandelbarkeit und Flexibilität zu. Dies ist auf die sensorischen Fähigkeiten von Menschen, insbesondere in Bezug auf ihr Greif- und Tastvermögen und ihre Fähigkeit, auf unvorhergesehene Situationen und Systemzustände zu reagieren, zurückzuführen. Schlussfolgernd werden Mensch und Technik vermehrt miteinander interagieren. Diese Überlegungen bilden die Grundlage der Social Networked Industry [1].

Das neuartige Zusammenspiel von Mensch und Technik sowie die Ausprägungen menschlicher Aktivitäten bedingen den Einsatz moderner Mess-, Darstellungs- und Auswertungsverfahren. Diese Verfahren umfassen Methoden der künstlichen Intelligenz (KI) zur Erkennung menschlicher Aktivitäten (eng. Human Activity Recognition, HAR). HAR-Methoden sind in der Lage, Sequenzen menschlicher Bewegungen, die mittels Sensoren erfasst und in eine maschinenlesbare Form gebracht werden, vordefinierten Aktivitäten zuzuordnen. Bei dieser Klassifikation handelt sich also um überwachtes Lernen. Der Hauptvorteil dieses Ansatzes besteht in seiner Automatisier- und Skalierbarkeit. Anders als klassische Methoden wie die REFA-Zeitstudie, bei denen Arbeitsabläufe manuell erfasst werden, können Bewegungen automatisch erfasst und wiedererkannt werden. Dabei beschränkt sich die Erkennung nicht auf einfache Aktivitäten wie *Stehen*, *Gehen* oder *Rennen*, sondern umfasst komplexe Anwendungsfelder aus der Produktion und Logistik.

HAR hat bereits in unserem Alltag Einzug gehalten. Ein Beispiel dafür sind Fitness-Tracker. Sie erfassen Bewegungssequenzen mittels inertialer Messeinheiten (IMU) und werten diese in Echtzeit aus. Als Ausgabe erhalten wir als Nutzer u. a. unsere Schrittzahl und Analysen zum Schlafverhalten. Darüber hinaus kann mittels Fitness-Tracker die ausgeübte Sportart erkannt werden. Neben dem alltäglichen Gebrauch ist HAR in weiteren Domänen der Wissenschaft und Industrie zu finden [2]: Im Gesundheitswesen beispielsweise wird das Pflegepersonal automatisch über Stürze von Patienten informiert [3]. Ein weiteres Anwendungsfeld umfasst Handbewegungen bzw. Gesten. Ihre Erkennung fördert neben der Mensch-Maschine-Interaktion [4] auch die Kommunikation im Rahmen der Gebärdensprache. Mittels HAR werden Handbewegungen automatisch zu Buchstaben, Wörtern und ganzen Sätzen übersetzt. Von weiteren zahlreichen Anwendungsfeldern sind speziell die Produktion [5] und Dienstleistungen in der Intralogistik [6,7] hervorzuheben. Im Vergleich zur HAR-Anwendung beim Fitness-Tracker, welcher zwischen *Schritt* und *kein Schritt* unterscheidet, sind die zu analysierenden Bewegungen in der Produktion und Intralogistik deutlich komplexer und die Übergänge zwischen verschiedenen Aktivitäten deutlich schneller getaktet.

Je komplexer die Bewegungen sind und je detaillierter diese analysiert werden sollen, desto herausfordernder wird die automatische Erkennung von Aktivitäten. Um dennoch eine hohe Performanz von HAR zu erreichen, können die Methoden um zusätzliche Datenströme erweitert werden – den sogenannten Kontext [8]. Der Kontext umfasst Angaben zu den Entitäten Mensch, Objekt und Hilfsmittel, die nicht unmittelbar die menschlichen Bewegungen umfassen [9]. Hierzu zählen u. a. ihr Zustand, ihre Identität oder ihre Position [10]. Steht beispielsweise eine Arbeitskraft neben einem Regal (Objekt) und hat einen Scanner (Hilfsmittel) in der Hand, dienen diese Daten einer verbesserten Aktivitätserkennung.

Damit HAR-Methoden Bewegungen aus Sensordaten erkennen können, müssen diese zuvor trainiert werden. Mittels annotierter Daten, in denen die Bewegungen bereits definierten Aktivitäten zugeordnet sind (Labeln), lernen HAR-Methoden verschiedene Muster in Bewegungs- und Kontextdaten wiederzuerkennen. Anschließend können unbekannte Bewegungen anhand dieser Muster definierten Aktivitäten zugeordnet werden. Die benötigten Daten werden im Innovationslabor des Lehrstuhls für Förder- und Lagerwesen (FLW) der Technischen Universität Dortmund entwickelt und in Kooperation mit Industriepartnern auf den industriellen Einsatz untersucht. Das Ziel von Aufnahmen in einem solchen Reallabor ist es, intralogistische Systeme bereits in der Planungsphase zu untersuchen und Daten zu erfassen. Mittels dieser Daten werden KI-Methoden trainiert, bevor ein System in Betrieb genommen wird.

## 2. Reallabor zur Bewegungsanalyse und Datenaufnahme

Das Reallabor am FLW ist ein Testraum für innovative Technologien, der ihre Erprobung unter realitätsnahen Bedingungen zulässt. Hier können neuartige Formen des Zusammenspiels von Mensch und Technik im Rahmen moderner Logistiksysteme erprobt werden. Ihr physisch erlebbares Abbild in einem Reallabor stellt aus vielerlei Gründen den bevorzugten Ausgangspunkt vieler Forschungsansätze dar.

Zunächst lassen sich in einem Reallabor auch solche Szenarien gefährdungsfrei abbilden, die in der Entwicklungsphase für Menschen potenziell sicherheitskritische Elemente aufweisen. Beispielsweise ist die Interaktion von Arbeitskräften mit Transportdrohnen, die sie über diverse Sensorschnittstellen autonom wahrnehmen sollen, mit Risiken durch Kollisionen verbunden. In einem Reallabor können die notwendigen Sicherheitsvorkehrungen getroffen und zusätzliche Sensorsysteme zur Überwachung eingesetzt werden, die im industriellen Realsystem wie beispielsweise einem Lager nicht zur Verfügung stünden. Diese Referenzsensoren der Laborumgebung ermöglichen eine zielgerichtete Verbesserung der zu entwickelnden Lösungen hin zur Praxistauglichkeit. Der Reallabor-Ansatz ist grundsätzlich nicht vom Vorhandensein eines real existierenden, in Betrieb befindlichen Logistiksystems abhängig. Somit ist eine realitätsnahe Erprobung der durch technologische Innovationen geänderten Prozessabläufe bereits in ihrer Planungsphase möglich. Beispielsweise können Sensoren und Klassifikatoren zur Aktivitätserkennung sogar für solche Logistiksysteme entwickelt werden, die zum Zeitpunkt der Laborexperimente noch nicht in Betrieb gegangen sind. Maßgeschneiderte Lösungen der menschlichen Aktivitätserkennung, die manuelle Aktivitäten und Prozesse in einem neu konzipierten oder umgestalteten Logistiksystem zeitlich und quantitativ erfassen sollen, stehen somit von Tag eins an zur Verfügung. Außerdem ermöglichen die Experimente und Aufnahmen in der Laborumgebung bereits Rückschlüsse auf Verbesserungspotenziale für das Realsystem.

Eine Laborumgebung liefert die Möglichkeit, verschiedene Sensortechnologien hinsichtlich ihrer Praxistauglichkeit für den jeweiligen Anwendungsfall vergleichend zu benchmarken. Für die Erfassung menschlicher Bewegungen werden im Reallabor u. a. IMUs, Videokameras und ein markerbasiertes Motion Capturing (MoCap) System verwendet. Das MoCap-System erfasst am Körper getragene und an Objekten befestigte Marker mittels Infrarotkameras. Aufgrund seiner hohen Genauigkeit dient das System als Referenz. Unpräzisere, aber für den industriellen Einsatz taugliche Sensoren werden durch diese Referenzumgebung untersucht. Für die Positionsbestimmung von Arbeitskräften, Objekten und Hilfsmitteln kommen beispielsweise Radio Frequency Identification, Ultra-Wideband, Bluetooth Low Energy, WLAN oder Indoor Global Positioning System in Frage. Mittels des MoCap-Systems kann für jede Technologie die optimale Anbringung der Sensoren ermittelt werden, die zu möglichst aussagekräftigen Kontextdaten führt.

## 3. Laufende Forschungsarbeiten im Reallabor

In verschiedenen Forschungsprojekten zur menschlichen Aktivitätserkennung sowie der Kontexterkennung kommt das Reallabor erfolgreich zum Einsatz. Folgender Überblick stellt ihre jeweilige Vision und insbesondere Anknüpfungspunkte für industrielle Transferprojekte heraus.

### 3.1. Attributbasierte Repräsentation von Klassen

Bisherige Arbeiten zur Aktivitätserkennung fassen Tätigkeiten wie beispielsweise *Fortbewegung* oder *Entnahme* jeweils als von einem Klassifikator automatisch wiederzuerkennende, entsprechend bezeichnete Aktivitätsklasse auf. Die zugrundeliegende Annahme, dass sämtliche Aktivitätsklassen im Vorhinein bekannt sind und sich ihre Anzahl, Definition und Abgrenzung im Nachhinein nicht verändern, ist in Zeiten sich intelligent an die Gegebenheiten anpassender Materialflusssysteme zunehmend kritisch zu sehen. Die Berücksichtigung jeweils neuer Klassen für jede neue Aktivitätsvariante verkompliziert die Annotation, d. h. das Setzen der Label an den aufgenommen Daten, mittels derer der Klassifikator trainiert wird und verringert die Beispielmenge je Klasse. Eine unmittelbare Assoziation eines Sensormusters zu einer spezifischen Aktivität ist nicht zielführend, um die Vielfältigkeit menschlicher Tätigkeiten abzubilden.

Einen Ausweg aus dem Dilemma zwischen einer möglichst hohen Detaillierung der Aktivitätsdefinition einerseits und ihrer dadurch sinkenden Übertragbarkeit zwischen verschiedenen Industrieanwendungen andererseits bieten sogenannte semantische Attribute. Dieser Ansatz stammt aus der Bilderkennung [11]. Forscher labelten Bilder von Tieren nicht allein mit ihrer Klasse, sondern mit einer semantischen Beschreibung, z. B. ob es sich um ein weißes oder schwarzes Tier handelt, in welchem Habitat es lebt usw. Mit diesen Attributen ist ein Klassifikator in der Lage, unbekannte Konzepte bzw. Klassen (in diesem Beispiel also Tiere) anhand ihrer semantischen Beschreibung zu erkennen. Übertragen auf menschliche Aktivitäten ist die Idee, Aktivitätsklassen durch semantische Beschreibungen zu repräsentieren. Sie sind im übertragenen Sinne Buchstaben, aus denen Wörter, d. h. Aktivitätsklassen, flexibel gebildet werden können. In [12] werden Attribute wie *Stehen*, *Laufen*, *Handhabung überkopf*, *Handhabung in neutraler Haltung*, *Linke Hand*, *Rechte Hand*, sowie *Posen* für Gegenstände wie *Werkzeuge* oder einen *Wagen* unterschieden. Ihre Kombination erlaubt die eindeutige Beschreibung beliebig definierter Klassen. Für weitere Informationen sei auf die Ausführungen in [13–18] verwiesen.

Dieser Ansatz ist mit der MotionMiners© GmbH bereits erfolgreich validiert worden. Im Reallabor aufgenommene Bewegungsdaten konnten für das Training eines Klassifikators verwendet werden, der eine vergleichbare Leistung bei der Aktivitätserkennung erzielt wie ein auf Daten aus den Realsystemen trainierter Klassifikator. Die heterogenen Aktivitätsdefinitionen in den verschiedenen Systemen und den Labordaten konnten mittels der semantischen Attribute ohne zusätzlichen Annotationsaufwand überbrückt werden.

### 3.2. Kontextbasierte Aktivitätserkennung

Wie im vorherigen Unterkapitel beschrieben können aus den semantischen Beschreibungen (Attribute) Wörter (Aktivitätsklassen) gebildet werden. Doch Wörter alleine haben eine begrenzte Aussagekraft, sie ergeben erst im Satz einen Sinn. Schlussfolgernd müssen die Aktivitätsklassen in einen Zusammenhang gebracht werden, damit die Informationen für Menschen verständlich sind. Durch weitere Datenströme, dem Kontext, können die erkannten Attribute mit u. a. der Identität und der Position von Menschen, Objekten sowie Hilfsmitteln und darüber hinaus mit Prozesswissen, Zustands- und Übergangslogik erweitert und durch einen kontextbasierten Klassifikator Aktivitätsklassen zugeordnet werden.

Das Ziel der Kontextverwendung ist die Leistungssteigerung der Aktivitätserkennung, indem die Kontextabhängigkeit der Aktivitäten genutzt wird: Ähnliche Bewegungen können unterschiedlichen Aktivitäten entsprechen, je nachdem, welches Objekt oder Hilfsmittel gehandhabt wird oder wo sich die Person befindet. Dabei baut der Ansatz auf der attributbasierten Repräsentation von Klassen auf, kann aber auch mit anderen Methoden und Klassifikatoren kombiniert werden.

Aus der Synchronisation von Aktivitätsklasse und Kontext ergibt sich schließlich ein verständlicher Satz. Dieser Zusammenhang wird anhand von Abbildung 1 verdeutlicht: Die Aktivitätsklasse und Attribute geben Aufschluss darüber, dass eine Person mit beiden Händen überkopf arbeitet (b). Erst durch die Hinzunahme des Kontextes (c) wird ersichtlich, in welchem Rahmen die Aktivität ausgeführt wird. Die Positionen zeigen, in welcher Gasse und vor welchem Regal sich die Person befindet. Werden Identitäten und Positionen kombiniert, wird deutlich, dass sich eine Kommissioniererin vor einem Kommissionierwagen befindet und einen Knopf am Put-to-Light Rahmen betätigt. Diese Kombination von Datenströmen kann verwendet werden, um die zuvor bestimmte Aktivitätsklasse zu überprüfen und ggf. zu korrigieren.

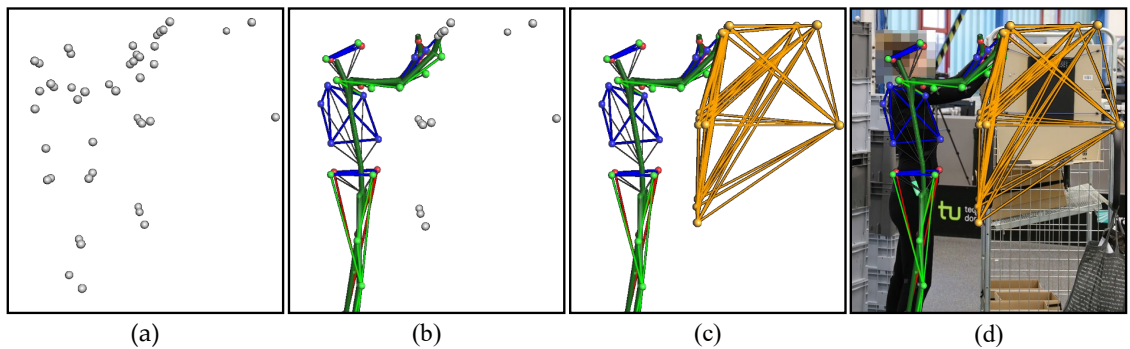

**Abbildung 1**: Datenverarbeitungsprozess des MoCap-Systems basierend auf einer Person-zu-Ware Kommissionierung: (a) Rekonstruktion einer Punktwolke bestehend aus reflektierenden Markern, (b) Rekonstruktion des Probanden aufgrund dessen Markermuster, (c) Rekonstruktion des Kommissionierwagens, (d) Überlagerung der MoCap-Visualisierung mit der Videoaufzeichnung [19].

Aus dem Prozesswissen leitet sich für die in Abbildung 1 visualisierte Bewegung der Prozess *Kommissionierung* sowie der Teilprozess *Bestätigung des Abgabevorgangs* ab. Die Ausführungsdauer des Teilprozesses kann aus den Daten erhoben werden. Durch die Zustands- und Übergangslogik lassen sich sogar vorherige Teilprozesse bestimmen und Aktivitätsübergänge plausibilisieren. Vor der Abgabebestätigung sollten eine oder mehrere Einheiten eines Artikels gepickt und in einem oder mehreren Behältern des Kommissionierwagens gelegt worden sein. Weicht das Ergebnis vom vorherigen Teilprozess ab, kann auch in diesem Schritt eine Anpassung der Aktivitätsklasse erfolgen. Durch Einschränkungen in der Übergangslogik kann eine grobe Schätzung des an die Abgabebestätigung anschließenden Teilprozesses getroffen werden. Als nächster Teilprozess ist die Entnahme von Einheiten einer neuen, nahegelegenen Position oder die Fortbewegung mit dem Kommissionierwagen zur nächsten Position bzw. zur Konsolidierung wahrscheinlich.

Mittels des im Reallabor erfassten Kontextes konnte die Leistung der Aktivitätserkennung erheblich gesteigert werden [20]. Zusätzlich wurden einzelne Kontextdaten hinsichtlich ihrer Relevanz für die Aktivitätserkennung analysiert [19]. Dies ermöglicht es, die Daten im Realsystem gezielt mit möglichst wenig Aufwand und zu/mit geringen Kosten zu erfassen. Die Erkenntnisse aus Aktivitätsklassen, Attributen und Kontextdaten können u. a. für Prozessanalysen und den cyberphysischen Zwilling verwendet werden.

### 3.3. Generierung menschlicher Bewegungsdaten - Cyberphysischer Zwilling

Die Erstellung eines Trainingsdatensatzes ist einem hohen Aufwand verbunden. Daten müssen aufgenommen, annotiert und anschließend manuell revidiert werden. Zudem steigt das notwendige Datenvolumen bei Zunahme komplexer Bewegungen, wie es beispielsweise in der Produktion oder Intralogistik der Fall ist. Als Folge werden HAR-Methoden überwiegend an einfachen Alltagssituationen getestet, die mit wenig Aufwand für die Vorbereitung und Durchführung von Aufnahmen einhergehen. Um zukünftig den Aufwand für die Datenaufnahme komplexer Umgebungen, wie der Intralogistik, zu vermeiden, werden menschliche Bewegungen simulationsbasiert generiert (siehe Abbildung 2). Anhand von Sensordaten aus der Intralogistik und anderen Domänen werden die menschlichen Bewegungen modifiziert, wodurch synthetische und damit neue Bewegungen entstehen [21]. Die Simulationsumgebung ermöglicht es, die synthetischen Bewegungen als unterschiedlichen Sensoroutput (z. B. IMU oder MoCap) zu exportieren. Je nach Datengrundlage kann die Modifikation sogar das Alter, Geschlecht, körperliche Einschränkungen oder Ermüdungserscheinungen der menschlichen Probanden simulativ berücksichtigen. Anhand dieser Variationen lassen sich personen- und situationsspezifische Simulationen durchführen, die realitätsnahe Bewegungsmuster entstehen lassen.

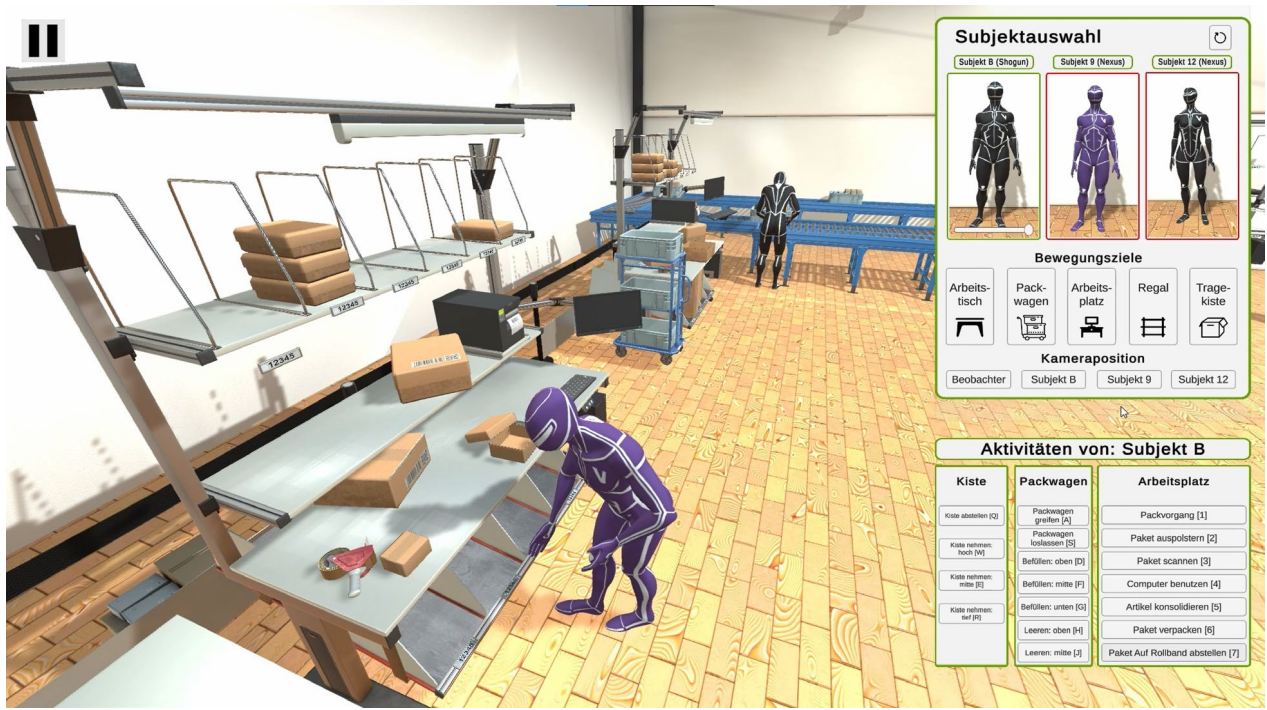

**Abbildung 2**: Visualisierung von Avataren in Unity während eines Verpackungsprozesses [22].

Der Ansatz der Datengenerierung ermöglicht die Überführung geplanter Systeme in eine Simulationsumgebung die Erzeugung synthetischer Bewegungsdaten zum Trainieren eines Klassifikators– und das, ohne das System in der realen Umwelt oder im Labor kosten- und zeitaufwendig zu implementieren.

### 3.4. Identitätserkennung und ihr Transfer

Das Bewegungsverhalten von Menschen ist nicht deterministisch. Selbst wenn dieselbe Person dieselbe Aktivität wiederholt ausübt, wird sie das Muster in den Sensordaten nicht exakt replizieren können. Vielmehr handelt es sich um „Wiederholungen ohne Wiederholung" [23]. Noch stärker werden die Abweichungen, wenn verschiedene Personen Aktivitäten ausführen. Trotz dieser auftretenden Abweichungen im Gesamten ist die Art und Weise eines Einzelnen, Aktivitäten auszuführen, so spezifisch, dass Identitäten in Sensordaten wiedererkannt werden können – ohne dabei auf visuelle Merkmale angewiesen zu sein [24]. Diese inhärenten Merkmale, an denen sich Individuen unterscheiden lassen, werden Bewegungsidiosynkrasien genannt. Es ist sowohl möglich, spezifische Identitäten zu unterscheiden, als auch Rückschlüsse auf körperliche Merkmale wie Körpermasse, Größe und Händigkeit zu ziehen. Die Entkopplung der personenspezifischen Bewegungsidiosynkrasien von den aktivitätsabhängigen Bewegungsprofilen bietet zum einen die Möglichkeit, die Identität eines Menschen in den Daten zu neutralisieren. Zum anderen können die Idiosynkrasien zwischen Aufnahmen transferiert werden. Eine Aktivitätsaufnahme einer Person A kann somit in einer Weise verändert werden, dass sie so aussieht, als hätte Person B sie ausgeführt. Dazu kommen Methoden des sogenannten Style Transfers zum Einsatz, die z. B. von DeepFake-Videos bekannt sind.

Für die industrielle Anwendung ergibt sich hieraus das Potenzial, die Identitäten von Individuen aus den Aufnahmen zu neutralisieren, während die Aktivitätserkennung weiterhin möglich bleibt. Zum anderen ermöglicht der Transfer von Idiosynkrasien die Anreicherung von Datensätzen durch mehr Variabilität in der Bewegungsausführung. Der auf diesen angereicherten Daten trainierte Klassifikator ist robuster gegenüber diversifizierten, körperlichen Merkmalen der aufgenommenen Personen.

### 4. Fazit – Das Reallabor als Schnittstelle zwischen Industrie und Forschung

Der Abschluss realwissenschaftlicher Forschungsarbeit ist die empirische Validierung im Rahmen ihrer Anwendung. Das Reallabor mit seinen hochgenauen Messinstrumenten dient dabei als ideale Referenzumgebung. Hier werden neue Methoden, Technologien und Software auf ihre industrielle

Tauglichkeit untersucht, ohne den laufenden Betrieb im Unternehmen zu beeinflussen. Dabei liegen insbesondere die Verknüpfung zwischen Technologien und neuen Methoden und die daraus resultierenden Auswirkungen auf das Gesamtsystem im Fokus. Speziell wurden Daten verschiedener Sensortechnologien, wie das MoCap-System und IMUs mit Simulationssoftware sowie HAR-Methoden verknüpft. Die in diesem Beitrag vorgestellten Forschungsarbeiten zeigten auf, wie Mensch und Technik in Zukunft miteinander interagieren können. Alle Arbeiten wurden bzw. werden in Kooperation mit verschiedenen Industriepartnern im realen Lager validiert. Die positiven Erfahrungen haben das Reallabor als Schnittstelle zwischen Industrie und Forschung bestätigt.

## Literatur